\def\aj{{AJ}}
\def\araa{{ARA\&A}}

\def\apj{{ApJ}}
\def\apjs{{ApJS}}

\def\mnras{{MNRAS}}
\def\nature{{Nature}}

\newcommand{\lya}{Ly$\alpha$} 
\newcommand{\lle}{L/L_{\rm Edd}} 
\newcommand{\mmin}{M_{\rm min}}
\newcommand{\beff}{b_{\rm eff}}
\newcommand{\hmpc}{h^{-1}{\rm Mpc}}
\newcommand{\tq}{t_Q}

\newcommand{\eg}{{\rm e.g.}}
\newcommand{\ie}{{\rm i.e.}}
\newcommand{\etal}{{\rm et al.}}

\newcommand{\msun}{M_{\odot}}

\def\farcm{\hbox{$.\!\!^{\prime}$}}

\documentclass[cup5b]{caps}
\usepackage{graphicx}
\usepackage{amssymb}
\usepackage{ociwsymp1}   

\HeadText{P. Martini}

\begin{document}

\pagenumbering{arabic}

\author[]{PAUL MARTINI\\The Observatories of the Carnegie Institution of 
Washington}

\chapter{QSO Lifetimes}

\begin{abstract}

\vspace{0.1in}

The QSO lifetime $\tq$ is one of the most fundamental quantities for 
understanding black hole and QSO evolution, yet it remains uncertain 
by several orders of magnitude. If $\tq$ is long, then only a small 
fraction of galaxies went through a luminous QSO phase. In contrast, 
a short lifetime would require most galaxies today to have undergone a 
QSO phase in their youth. The current best estimates or constraints on 
$\tq$ from black hole demographics and the radiative properties of QSOs 
vary from at least $10^6$ to $10^8$ years. This broad range still allows 
both possibilities: that QSOs were either a rare or a common stage of 
galaxy evolution. These constraints also do not rule out the possibility 
that QSO activity is episodic, with individual active periods 
much shorter than the total active lifetime. 

In the next few years a variety of additional observational constraints 
on the lifetimes of QSOs will become available, including clustering 
measurements and the proximity effect. These new constraints 
can potentially determine $\tq$ to within a factor of 3 and therefore 
answer one of the most fundamental questions in black hole evolution: Do 
they shine as they grow? This precision will also test the viability of 
our current model for accretion physics, specifically the 
radiative efficiency and need for super-Eddington 
luminosities to explain the black hole population. 

\end{abstract}

\section{Introduction}

\vspace{0.1in}

Shortly after the discovery of luminous QSOs, several authors suggested 
that they could be powered by accretion onto supermassive black holes
(\eg, Salpeter 1964; Zel'dovich \& Novikov 1964; Lynden-Bell 1969). 
The strong evidence that dormant, supermassive black holes exist at the 
centers of all galaxies with a spheroid component (\eg, Richstone \etal\ 1998) 
supports this hypothesis and suggests these galaxies go through an optically 
luminous accretion phase. More recent observations suggest that the masses of 
supermassive black holes are correlated with the sizes of their host galaxy 
spheroids, which implies some connection between the growth of the 
subparsec-scale supermassive black holes and the kiloparsec-scale spheroids. 
The critical parameter for understanding how quickly these present-day black 
holes grew to their present size is the lifetime of the QSO phase $\tq$. 
The QSO lifetime determines the net growth of supermassive black holes 
during an optically luminous phase. It may also have implications for 
the growth rate of galactic spheroids. 

For the purpose of the present review, the lifetime of a QSO is defined to be 
the total amount of time that accretion onto a supermassive black hole is 
sufficiently luminous to be classified as a QSO --- that is, the net time 
the luminosity in some wavelength range $L > L_{Q,min}$, where $L_{Q,min}$ is 
the minimum luminosity of a QSO. 
This is an observational definition, chosen because QSOs are categorized on 
the basis of luminosity in existing surveys. A more physically motivated 
definition of a QSO would define the lifetime as the time a black hole 
is accreting above some minimum mass accretion rate, or emitting radiation 
above some minimum fraction of the Eddington luminosity (and this definition 
would also require a minimum black hole mass), as these two parameters are 
more directly relevant to the details of the accretion physics. The measured 
QSO luminosity instead depends on the product of one of these quantities with 
the black hole mass.  

A valuable fiducial time scale for supermassive black hole growth is the 
Salpeter (1964) or $e$-folding time scale: 
\begin{equation}
t_S = M/\dot{M} = 4.5 \, \times \, 10^7 \, \left(\frac{\epsilon}{0.1}\right) \, 
\left(\frac{L}{L_{\rm Edd}}\right) \, {\rm yr} , 
\end{equation}
where $\epsilon = L/\dot{M} c^2$ is the radiative efficiency for a 
QSO radiating at a fraction $\lle$ of the Eddington luminosity.
Commonly accepted values of these two key parameters for luminous QSOs 
are $\epsilon = 0.1$ and $\lle = 1$. 
The importance of an optically luminous or QSO mode of black hole growth 
relative to significantly less efficient, or less luminous modes is not well 
established, primarily due to uncertainties in these two parameters and the 
uncertainty in the QSO lifetime. If the QSO lifetime is 
long, then QSOs were exceptionally rare objects and only a small number 
of the present-day supermassive black holes went through a QSO phase. 
In this scenario, essentially all of the mass in these black holes can be 
accounted for by luminous accretion. The remaining supermassive black holes 
in other galaxies must instead have accreted their mass through less 
radiatively efficient or less luminous modes. If instead the QSO lifetime 
is short, most present-day black holes went through an optically luminous 
accretion phase, yet this phase made a relatively small contribution to the 
present masses of the black holes. 

One long-standing upper limit on the lifetime of QSOs is the lifetime of the 
entire QSO population. The evolution of the QSO space density is observed 
to rise and fall on approximately a $10^9$ year time scale 
(see, \eg, Osmer 2003).  More refined demographic arguments, 
described in detail in the next section, find values on order 
$10^6 - 10^7$ years. Lower limits to the QSO lifetime are less direct, such 
as that provided by the proximity effect in the \lya\ forest,  
and they argue for a lower limit of $\tq > 10^4$ years. 
Unlike demographic arguments, which constrain the net time a black 
hole accretes matter as a luminous QSO, the lower limits are based on 
radiative measures that do not take into account the possibility that 
a given black hole may go through multiple luminous accretion phases. 
These radiative methods, described in \S\ref{sec:epi}, constrain the 
episodic lifetime of QSOs and place a lower bound on the net lifetime. 

Observations of QSOs and the less-luminous Seyfert and LINER galaxies 
demonstrate that AGNs have a range of different accretion mechanisms 
and these mechanisms depend on AGN luminosity. 
In the present review I will only discuss the luminous QSOs, defined to  
be objects with absolute $B$ magnitude $M_B > -23$ mag. These objects 
may form a relatively homogeneous subclass of AGNs that 
accrete at an approximately fixed fraction of the Eddington rate 
with constant radiative efficiency. This luminosity limit corresponds to 
$L \approx L_{\rm Edd}$ for a $10^7\, \msun$ black hole. 

\section{The Net Lifetime} \label{sec:net}

\vspace{0.1in}

The main methods used to estimate the net lifetime of QSOs are demographic 
estimates, which typically take one of two forms: integral or counting. 
Integral estimates are based on the integrated properties of QSOs over 
the age of the Universe, while counting arguments compare the numbers of 
objects at different epochs. The classic integral constraint on $\tq$ 
is that the amount of matter accreted onto QSOs during their lifetime $\tq$, 
as represented by the luminosity density due to accretion, should be 
less than or equal to the space density of remnant black holes in the 
local Universe (So\l tan 1982). This type of estimate is 
most sensitive to the assumed value of the radiative efficiency $\epsilon$. 
One example of a counting argument is to ask what value of $\tq$ is 
require if all bright galaxies go through a QSO phase in their youth 
(\eg, Rees 1984). 
Counting arguments are not sensitive to $\epsilon$, but are affected 
by the assumed value of $\lle$ and galaxy (black hole) mergers. 

The predicted upper bound for $\epsilon$ is 0.3, attained in models of thin 
disk accretion onto a maximally rotating Kerr black hole (Thorne 1974). 
The value of $\epsilon$ is more commonly set to $0.1$, based on 
expectations for thin-disk accretion onto a non-rotating black hole. 
Determination of $\lle$ is more tractable observationally, at 
least once the mass of the central black hole is known. 
The most common assumption is that $\lle \sim 1$ for luminous QSOs. 
If they are substantially sub-Eddington, then this implies the presence of a 
population of massive ($> 10^{10}\, \msun$) black holes not seen in the 
Universe today. In contrast, 
substantial super-Eddington luminosities (greater than a factor of a few) 
have historically been considered unlikely due to the limitations of 
radiation pressure that define the Eddington limit. However, recent work has 
suggested that the Eddington luminosity could be exceeded by as much as a 
factor of 100 in some objects due to small-scale inhomogeneities in thin 
accretion disks (Begelman 2002). 

The demographic constraint set by the evolution of the entire QSO population, 
$\tq \leq 10^9$ years, is essentially the only demographic constraint on 
$\tq$ that does not depend on some assumptions about the parameters 
$\epsilon$ and $\lle$ and is neither a counting nor an integral constraint. 
If the conventional values for these parameters are assumed and $\tq = 10^9$ 
years, then the black holes in QSOs grow by $\sim 25$ $e$-folds during this QSO 
phase. This rapid, prolonged growth would produce a population of extremely 
massive black holes ($> 10^{10}\, \msun$) that have not been observed in the 
local Universe (Cavaliere \& Padovani 1988, 1989), assuming the initial 
black hole seeds are stellar remnants with $M > 1\, \msun$. 

A simple counting argument was considered by Richstone \etal\ (1998), 
who compared the space density of QSOs at $z \approx 3$ to supermassive black 
holes in the local Universe. As the space density of QSOs at their peak is 
approximately $10^{-3}$ times that of present-day supermassive black holes, 
the implied lifetime is $\tq \approx 10^6$ years. 
This lifetime is much shorter than the Salpeter time scale and therefore 
implies that only a small percentage of the present-day mass in black holes 
was accreted during the QSO epoch.

\subsection{QSO Evolution and the Local Black Hole Population}

\vspace{0.1in}

Many more detailed demographic models have been put forth, particularly in the 
last decade (Small \& Blandford 1992; Haehnelt \& Rees 1993; Haehnelt, 
Natarajan, \& Rees 1998; Salucci \etal\ 1999; Yu \& Tremaine 2002). These 
models are mostly based on counting arguments that simultaneously address the 
present-day black hole mass function and the space density evolution of QSOs 
with a recipe for the luminosity evolution and growth of supermassive black 
holes in the QSO phase. Haiman \& Loeb (1998) and Haehnelt \etal\ (1998) find 
that they can match the present-day black hole mass function and the QSO 
luminosity function at $z = 3$ with $\tq$ between $10^6$ and $10^8$ years, 
depending on the relationship between the mass of the central black hole and 
the host halo. They describe how a lifetime as short as $10^6$ years requires 
that much of the present-day black hole mass is not accreted as an 
optically luminous QSO. 

  \begin{figure}
    \centering
%    \vspace{1cm}
    \includegraphics[width=100mm]{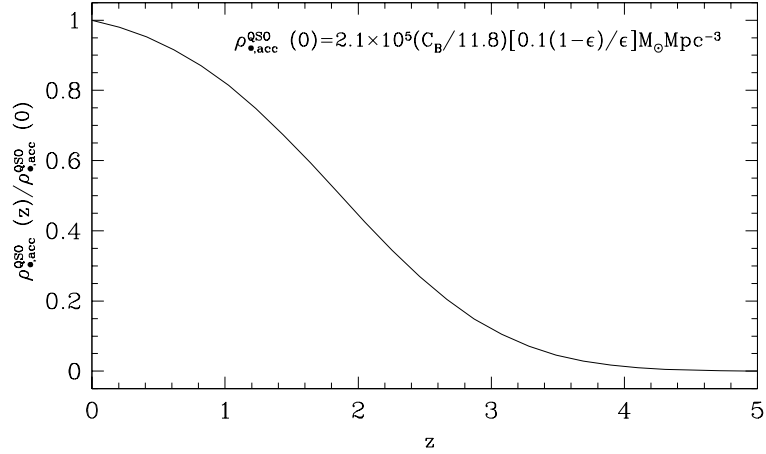}
    \caption{History of accretion onto supermassive black holes by 
optically luminous QSOs (from Yu \& Tremaine, their Fig.~1). }
    \label{fig:acc}
  \end{figure}

Yu \& Tremaine (2002; see also Yu 2003) develop a similar model that is 
a variation on So\l tan's (1982) integral approach 
using new data for the present-day black hole mass function, as well as 
the luminosity function and space density evolution of QSOs. 
They calculate the present-day black hole mass function with the 
$M_\bullet  - \sigma$ relation (Ferrarese \& Merrit 2000; Gebhardt \etal\ 2000) 
and velocity dispersion measurements for a large number of elliptical galaxies 
and massive spheroids from the SDSS (Bernardi \etal\ 2003) and take the 
QSO luminosity function from the 2dF (Boyle \etal\ 2000) and the Large Bright 
QSO Survey (Hewett, Foltz, \& Chaffee 1995). Their estimate of the current 
mass density in black holes is 
$\rho_\bullet^{\rm QSO} (z = 0) = (2.5 \pm 0.4) \times 10^5 h_{0.65}^{2}\, \msun {\rm Mpc}^3$. 
This can be compared to the integrated mass accretion onto black holes 
due to luminous QSOs, which they calculate to be
\begin{equation}
\rho_{\bullet,{\rm acc}}^{\rm QSO} (z) = \int_{z}^{\infty}\frac{dt}{dz}dz 
\int_0^{\infty} \frac{(1-\epsilon) L_{bol}}{\epsilon c^2} \Phi(L,z) dL, 
\end{equation}
where $\Phi(L,z)$ is the QSO luminosity function. 
The total mass density accreted by optically luminous QSOs is then 
$\rho_{\bullet,{\rm acc}}^{\rm QSO} (z = 0) = 2.1 \times 10^5 (C_B/11.8) 
[0.1(1-\epsilon)/\epsilon] \, \msun {\rm Mpc}^3$, 
where $C_B$ is the bolometric correction in the $B$ band scaled by the 
value of $11.8$ calculated by Elvis \etal\ (1994). The cumulative accretion 
history they derive is shown in Figure~\ref{fig:acc}. Note that half of the 
total mass density accreted by QSOs occurred by $z \approx 1.9$. 

  \begin{figure}
    \centering
%    \vspace{1cm}
    \includegraphics[width=100mm]{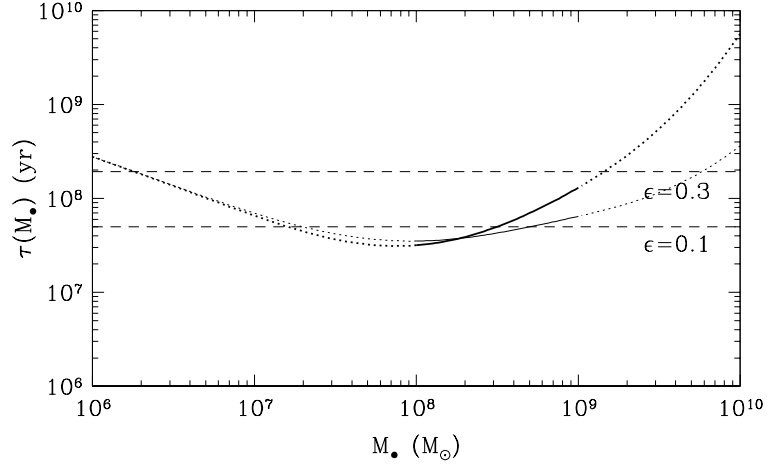}
    \caption{Mean QSO lifetime from Yu \& Tremaine (2002; their Fig.~5). 
The two lines represent the calculated lifetime for two values for the 
radiative efficiency $\epsilon$. The solid parts of the lines show the range 
of local black hole masses estimated from observations of early-type galaxies. 
The dotted lines represent regions where the local black hole mass function 
and QSO luminosity function were extrapolated. The two horizontal, dashed 
lines are the Salpeter time scale for $\epsilon = 0.3$ and 0.1. }
    \label{fig:tau}
  \end{figure}

%% paragraph on lifetimes 

The mean lifetime of QSOs above some black hole mass is then approximately 
equal to the integrated luminosity of accreted matter by these black holes,  
divided by their space density: 
\begin{equation}
\tau(>M_{\bullet}) \simeq \frac{\int^\infty_{L(M_{\bullet})} dL \int^{t_0}_0 \Phi(L,t) dt}{\int^\infty_{M_{\bullet}} n_{M_{\bullet}}^{\rm early} (M_{\bullet}', t_0) dM_{\bullet}' }. 
\end{equation}
This method is therefore essentially an integral method, but by 
considering the lifetime above a range of black hole masses Yu \& 
Tremaine (2002) capture some of the elements of a counting argument as well. 
Figure~\ref{fig:tau} shows the mean lifetime derived by Yu \& Tremaine 
(2002) as a function of black hole mass for two values of the radiative 
efficiency. The solid lines mark the range of black hole mass constrained 
by the velocity dispersions of early-type galaxies. The dotted lines are 
based on extrapolations of the local black hole mass function and 
the luminosity function of QSOs. This calculation assumes that QSOs are 
radiating at the Eddington luminosity and that the spectral energy 
distributions of the entire QSO population are well represented by the 
Elvis \etal\ (1994) sample. 
For $\epsilon = 0.1 - 0.3$ and $10^8 < M_{\bullet} < 10^9\, \msun$, 
Yu \& Tremaine calculate that the mean QSO lifetime lies in the range 
$\tq = 3 - 13 \times 10^7$ years. 

%% paragraph on backgrounds 

For the standard radiative efficiency of $\epsilon = 0.1$ and a bolometric 
correction $C_B = 11.8$, optically selected QSOs can account for 
$\sim 75$\% of the current matter density in black holes. 
The X-ray and infrared backgrounds offer a potentially better constraint on 
the total mass accretion onto black holes, as these wavelengths are less 
susceptible to obscured or highly beamed AGNs. Barger \etal\ (2001) combined 
these multiple backgrounds and estimated  
$\rho_{\bullet,{\rm acc}} \approx 9 \times 10^5 (0.1/\epsilon)\, \msun {\rm Mpc}^3$, 
higher than Yu \& Tremaine's estimate due to optically selected QSOs alone. 
The discrepancy may be due to a genuine population of obscured AGN, but then 
comparison of this higher accreted mass density with that estimated for local 
black holes implies $\epsilon \geq 0.3$, in conflict with the maximum 
efficiency estimated for accretion onto a Kerr black hole. Other potential 
causes of this discrepancy include (1) uncertainties in the local 
black hole mass density, which has been extrapolated to $M_{\bullet} < 
10^8\, \msun$ and $M_{\bullet} > 10^9\, \msun$, (2) a different population 
responsible for these 
backgrounds, (3) the possibility of a free-floating supermassive black hole 
population that has been ejected from galaxies, 
and (4) the possibility that a significant amount of black hole mass has been 
lost as gravitational radiation in mergers. 

\subsection{Coevolution of QSOs, Black Holes, and Galaxies}

\vspace{0.1in}

An alternative approach to modeling the coevolution of QSOs, black holes, 
and galaxies is to build detailed models with some adjustable parameters and 
ask what the lifetime is in that context. 
This is the approach adopted by Kauffmann \& Haehnelt (2000), who 
model the formation of QSOs during major mergers of gas-rich galaxies. 
Any merger between 
galaxies with a mass ratio greater than 0.3 results in some fraction 
$M_{acc}$ of gas accreted onto the black hole over a time scale $\tq$. 
The galaxy mergers and their cold gas content are then obtained from their 
semi-analytic model of galaxy evolution, combined with the merging 
history of dark matter halos from extended Press-Schechter (1974) theory.

Kauffmann \& Haehnelt let the peak luminosity of each QSO scale as the mass 
of gas accreted 
onto the central black hole, which in turn scales as the mass of available 
cold gas in the merger and the lifetime of the accretion phase. 
The radiative efficiency is taken to be constant and chosen so that the most 
luminous QSOs do not exceed the Eddington luminosity. The luminosities of QSOs 
after 
the merger do not remain constant, but decline exponentially from their peak 
on a time scale $\tq$. While their high-redshift QSOs emit at close to the 
Eddington luminosity, the QSOs at low redshift have $\lle = 0.01 - 0.1$. 
In addition to a QSO lifetime that is constant with 
redshift, these authors also parametrize the lifetime to scale as the dynamical 
time: $\tq(z) = \tq(0)\,(1 + z)^{-1.5}$, where $\tq(0)$ is the lifetime 
at $z = 0$. 
The lifetime that best fits the evolution of the QSO population, the 
merger history of galaxies, and the evolution of their cold gas fraction 
is $\tq(0) = 3 \times 10^7$ years (for a given merger event). 

\section{Limits on Episodic Activity} \label{sec:epi}

\vspace{0.1in}

\subsection{QSOs and Starbursts}

\vspace{0.1in}

The presence of many QSOs in merging systems and the detection 
of luminous hard X-ray emission from the cores of some merging 
galaxies strongly suggest that at least some QSOs are triggered 
by interactions of giant, gas-rich galaxies. The merging process 
removes angular momentum from a significant fraction of the two galaxies' 
ISM, leading it to flow inward toward the circumnuclear 
region and central black hole. This angular momentum loss is a natural 
mechanism to ignite QSO activity, as well as significant amounts of 
circumnuclear star formation. 

In extreme circumstances Norman \& Scoville (1988) proposed that mass loss 
from a sufficiently large star cluster formed in such a merger may provide a 
continuous fuel supply for a long $\tq$.  
These authors specifically explore the implications of a star cluster of mass 
$4 \times 10^9 M_{\odot}$ within the central 10 pc of the merger remnant. 
In $\sim 10^8$ years, mass loss from this cluster can power accretion 
rates of up to $10 M_{\odot}$ yr$^{-1}$, corresponding to QSO luminosities 
$L > 10^{12} L_{\odot}$ due to the black hole alone, and much higher than the 
stellar luminosity of the central star cluster. This model therefore 
predicts an episodic QSO lifetime of $\sim 10^8$ years after a major 
merger between two massive, gas-rich galaxies. 

A number of examples of so-called transition QSOs, which show evidence for 
recent and substantial star formation, were studied in detail by Canalizo 
\& Stockton (2001). They derived ages for the post-starburst population as high 
as 300 Myr in some objects, while Brotherton \etal\ (1999) derived an age of 
400~Myr for the post-starburst population in UN J1025-0040. If the the QSO and 
starburst were triggered at the same time, the implied lifetimes for the 
QSOs are greater than a few $\times 10^8$ years. 

Kawakatu, Umemura, \& Mori (2003) recently considered a similar coevolution 
model for black holes and galaxies. They model the formation and growth 
of the central black hole via radiation drag on the host galaxy ISM. This 
process gradually increases the mass of the central black hole for 
$\sim 10^8$ years, during which time the galaxy may appear as an ultraluminous 
infrared galaxy. 
Once the central region is no longer obscured, the galaxy appears as 
an optically identifiable QSO. This evolutionary scenario corresponds to 
a great deal of supermassive black hole growth, which these authors 
propose would lead to larger velocities in the broad-line region.

\subsection{Size of Narrow-Line Regions}

\vspace{0.1in}

The size of the narrow-line region (NLR) around AGNs potentially provides a 
straightforward, geometric estimate of their lifetime. 
Bennert \etal\ (2002) recently studied a sample of Seyferts and QSOs and found 
that the size of the NLR increases approximately as the square root of the 
[O~III] luminosity and the square root of an estimate of the H$\beta$ 
luminosity. 
For a constant ionization parameter, electron density, and covering 
factor, the size of the ionized region should scale as the square root 
of the luminosity. The NLR is therefore ionization bounded and the 
size of the ionized region sets a lower limit to the lifetime of the 
ionizing source. As the most luminous QSOs in their sample have NLRs  
approximately 10 kpc in radius, the episodic lifetime of the most luminous 
QSOs must be greater than $3 \times 10^4$ yr. 

\subsection{Lengths of Jets}

\vspace{0.1in}

Jets are another macroscopic product of accretion onto a supermassive black 
hole.  
Their expansion time therefore sets a lower limit to the lifetime of the 
accretion. The main uncertainty in the use of jets to constrain the 
lifetime is the expansion speed, although the lengths of the jets themselves 
are also somewhat uncertain due to the unknown inclination angle.  

The expansion speed of jets as a population can be estimated by the mean ratio 
of the length of the jet to the counterjet (Longair \& Riley 1979), although 
this requires that we assume that all jets have the same expansion speed. 
For an expansion speed $\upsilon = \beta c$ and inclination to the line of 
sight $\theta$, the observed length ratio of the jet to counterjet (following 
Scheuer 1995) is 
\begin{equation}
Q = (1 + \beta \, {\rm cos} \, \theta)/(1 - \beta \, {\rm cos} \, \theta). 
\end{equation} 
If we assume that the inclination angle of the population is uniformly 
distributed between zero and $\theta_{max}$ degrees, the mean of {\rm log} $Q$ 
in the limit of small $\beta$ is
\begin{equation}
\left<{\rm ln} \, Q\right> = \beta \, ( 1 + {\rm \cos} \, \theta_{max})[1 + \frac{1}{6} \, \beta^2 \, (1 + {\rm cos}^2 \, \theta_{max}) . . . ].
\end{equation}
Combining studies of radio jets by a number of authors, Scheuer (1995) 
obtains $\beta = 0.03 \pm 0.02$ and 
sets strong bounds on $\beta$ such that $0 < \beta < 0.15$. 

Blundell, Rawlings, \& Willott (1999) compiled data from three separate, 
flux-limited radio surveys and studied the evolution of the classical 
double population, using this large database to decouple redshift and 
luminosity degeneracies. 
With the expansion speed constraints of Scheuer (1995), they find 
ages for classical double sources as large as a few $\times 10^8$ years 
in the lowest redshift sample (such large sources are selected against 
at higher redshift due to the steepening of their spectra). The existence of 
these objects therefore appears to set a long lower limit to the lifetime 
of nuclear activity. 

The main caveat in the use of maximum jet lengths to place lower limits
on the episodic lifetime of QSOs is that jets are commonly found in very
low-luminosity AGNs, and not only in systems of QSO luminosities. While
nearly all of the sources used in Scheuer's study had QSO luminosities, 
this is not true of all of the sources studied by Blundell \etal\ (1999). 
The lifetime derived from the lengths of jets therefore 
do not exclude the possibility
that a high-luminosity QSO phase is a relatively short-lived stage in the 
lifetime of an AGN.

\subsection{Proximity Effect}

\vspace{0.1in}

QSOs can create a sphere of ionized matter in their vicinity whose radius 
is set by the episodic lifetime or recombination time scale, whichever is 
shorter. 
This ionized region is observed as the proximity effect in the \lya\ forest
(Bajtlik, Duncan, \& Ostriker 1988; Scott \etal\ 2000), 
which quantifies the decrease in the 
number of neutral hydrogen clouds of a given column density per unit redshift 
in the vicinity of the QSO. 
In order to produce this zone of increased ionization and decreased 
number density of the high-column density \lya\ forest clouds,
the ionization field from the QSO must have been operating for at least 
the recombination time scale of these relatively high-column density clouds. 
Because the proximity effect is observed along the line of sight to the QSO, 
the physical extent of the region only depends on the observed luminosity of 
the QSO, and not the episodic lifetime. 

The equilibration time scale for \lya\ clouds is: 
\begin{equation}
\tau \equiv n_{\rm HI} \left(\frac{dn_{\rm HI}}{dt}\right)^{-1},
\end{equation}
where $n_{\rm HI}$ is the number density of neutral hydrogen atoms. 
The relevant time scale is 
\begin{equation}
\tau = 1 \times 10^{4} (1 + \omega)^{-1} J_{21}^{-1} \ \ \ \ {\rm yr},
\end{equation}
where $\omega$ is the increase in ionizing flux due to the QSO relative 
to the background and $J_{21}$ the intensity of the background ionizing 
radiation at the Lyman-limit in units of $10^{-21}$ erg cm$^{-2}$ Hz$^{-1}$ 
sr$^{-1}$. The equilibration time scale is $\sim 10^4$ years for typical values 
of $\omega$ and $J_{21}$. 

\section{Future Prospects} \label{sec:fut}

\vspace{0.1in}

\subsection{Luminous QSOs at $z > 6$}

\vspace{0.1in}

Observation of QSOs in the early Universe when the optical depth in neutral 
hydrogen is large ($z > 6$) offers a more powerful constraint on the lifetime 
than the proximity effect because these QSOs exist in a neutral, predominantly 
lower-density IGM. 
Above $z \approx 6$, the Gunn-Peterson trough in the IGM will erase all flux 
blueward of \lya\ in the absence of ionization of the surrounding medium 
by the QSO, \ie\ an H~II region. The size of this region is set by 
the lifetime of the QSO as the sole ionizing source, unlike the 
case for the proximity effect, where the lifetime is only constrained to be 
the recombination time because only high density regions remain to be 
(re)ionized.  

Haiman \& Cen (2002; see also Haiman \& Loeb 2001) applied this concept to 
the $z = 6.28$ QSO 
SDSS 1030+0524. They used the expected density distribution of 
the IGM with a hydrodynamic simulation (Cen \& McDonald 2002) and 
found that a quasar lifetime of $2 \times 10^7$ years and an 
H~II region $\sim 4.5$ Mpc (proper) radius produced a good match to 
the observed flux in SDSS 1030+524 blueward of \lya. 

The masses of the black holes powering QSOs such as this one and 
others with $z > 6$ provide some additional information on the 
early evolution of QSOs. For example, one of the key conclusions 
of Haiman \& Cen (2002) was that the $z = 6.28$ QSO was not significantly 
lensed or beamed. Therefore if the QSO is shining at or below the 
Eddington rate, the mass of the central black hole is on order 
$2 \times 10^9 \, \msun$. The other $z > 6$ QSOs discovered by the 
SDSS have comparable luminosities, and therefore are presumably 
similarly massive. To form such a massive black hole at this early 
epoch requires that it has been accreting at the Eddington rate for 
nearly the lifetime of the Universe, less than $10^9$ years at $z = 6$ 
assuming $\epsilon = 0.1, L \approx L_{\rm Edd},$ and a $100 \, \msun$ seed. 
Depending on the mass of the seed black hole population, either the 
QSO lifetime must be quite long or super-Eddington accretion must occur 
in the early Universe. 

\subsection{Transverse Proximity Effect}

\vspace{0.1in}

The proximity effect can be used to set a stronger constraint on the episodic 
lifetime with a direct measure of its extent in the plane of the sky. 
This measurement could be made through observations of multiple QSOs that lie 
in close proximity on the sky. The proximity effect due to a lower-redshift 
QSO in the spectrum of higher-redshift QSO would provide a lower limit on 
the lifetime of the lower-redshift QSO equal to the time required to ionize the 
intervening neutral medium, where the ionization front expands at a speed 
$\upsilon \simeq c$ in the primarily low-density IGM. 
The angular diameter distance between the two QSOs is then a lower-limit 
to the lifetime. 
Such a measurement was first discussed by Bajtlik \etal\ (1988), 
although in practice the statistical significance of the proximity effect in 
neutral hydrogen of any given QSO at $z < 6$ is sufficiently weak that this 
measurement would require combining data from large numbers of QSOs to obtain 
a mean lower limit to the episodic lifetime.

The proximity effect in neutral hydrogen is difficult to measure 
because its mean opacity of the intergalactic medium is so low 
(at least below $z \approx 6$) and the recombination times for the remaining 
neutral, high-density clouds are quite short. 
Helium offers a better chance of success as the mean opacity of neutral 
helium is quite high above $z \approx 2.9$. Individual helium ionizing sources, 
\ie\ QSOs, above this redshift will therefore produce distinct opacity gaps.  
Unfortunately, relatively few He~II opacity gaps are known as the He~II \lya\ 
$\lambda 304$ \AA\ line is still in the space UV regime at $z \approx 3$. 
Current instrumentation therefore requires them to be observed in the 
spectra of very bright QSOs. These systems are also easily obscured by the 
much more common Lyman-limit systems. 

There are currently five known helium absorption systems in the spectra of 
four QSOs. These systems offer potentially excellent lower limits on the 
episodic lifetime if the origins of these systems are identified with QSOs 
a significant distance from the He~II system. Jakobsen \etal\ (2003) 
identified such a candidate QSO for the He~II opacity gap 
at $z \approx 3.056$ in the spectrum of the $z = 3.256$ QSO 
Q0302-003. They identified a $z = 3.050 \pm 0.003$ QSO at the 
approximate redshift of the helium feature with an angular 
separation of 6\farcm5 from the line of sight to Q0302-003. 
If the new QSO is responsible for the helium opacity gap, then the light 
travel time sets a lower limit of $\tq > 10^7$ yr. 

The widths of all of the known He~II opacity gaps are 
$\Delta z \simeq 0.01 - 0.02$, which corresponds to a size of $2 - 5$ Mpc 
(Jakobsen \etal\ 2003). These sizes set a strong lower limit of 
$\sim 10^6$ years for the lifetime of the ionizing source. An 
alternative explanation of these opacity gaps is that 
they are due to low-density regions in the IGM. Heap \etal\ (2000) 
studied the opacity gap in Q0302-003 and argue against this interpretation,  
as their simulations cannot reproduce the distribution and amplitude of the 
gap with a low-density region.  

\subsection{Clustering}

\vspace{0.1in}

The clustering of QSOs can provide a strong constraint on their net lifetime ,
under the assumption that QSOs reside in the most massive dark matter halos. 
If QSOs are long-lived, then they need only reside in the most massive, 
and therefore most strongly clustered, halos in order to match the observed 
space density of luminous QSOs. In contrast, if QSOs are short-lived 
phenomena, then a much larger population of host halos is required to 
match the observed QSO space density at the epoch of observation. As 
this large population of hosts will be dominated by less massive halos, 
due to the shape of the halo mass function, 
the QSO population will not be as strongly clustered. These simple 
arguments therefore predict that the larger the observed correlation 
length, the longer the lifetime of QSOs (Haehnelt \etal\ 1998). 

Martini \& Weinberg (2001) calculated the expected relation between QSO 
clustering and lifetime in detail in anticipation of forthcoming measurements 
from the 2dF and SDSS collaborations (see also Haiman 
\& Hui 2001). The main assumptions of this model are that the luminosity 
of a QSO is a monotonic function of the mass of its host dark matter halo 
and that all sufficiently massive halos go through a QSO phase. For some 
absolute magnitude-limited sample of QSOs at redshift $z$, the probability 
that a given halo currently hosts a QSO is simply $\tq/t_H$, where $t_H$ is 
the lifetime of the host halo. The QSO therefore does not necessarily turn 
on when the halo forms, which differs slightly from the model developed 
by Haehnelt \etal\ (1998). 

Calculation of the clustering corresponding to some lifetime $\tq$ requires 
knowledge of the minimum halo mass, which can be found by setting the 
integral of the mass function of dark matter halos, multiplied by the fraction 
that host QSOs, equal to the observed space density $\Phi(z)$: 
\begin{equation}
\Phi(z) = \int_{M_{\rm min}}^\infty dM {\tq \over t_H(M,z)} n(M,z).
\label{eqn:phimatch2}
\end{equation}
The halo lifetime $t_H$ depends on both the halo mass and redshift and was 
calculated by solving for the median time a halo of mass $M$ at redshift 
$z$ will survive until it is incorporated into a new halo of mass $2M$. 
The mass function of dark matter halos $n(M,z)$ is calculated with the 
Press-Schechter formalism: 
\begin{equation}
n(M, z)\;dM = - \sqrt{\frac{2}{\pi}} \frac{\rho_0}{M}
                \frac{\delta_c(z)}{\sigma^2(M)} \frac{d\sigma(M)}{dM} \;
                {\rm exp} \left[ -\frac{\delta_c^2(z)}{2 \sigma^2(M)} \right]
                dM .
\label{eqn:ps}
\end{equation}
The predicted clustering of QSOs depends directly on the minimum halo mass. 
The bias factor for halos of a given mass was calculated by Mo \& White 
(1996) as 
\begin{equation}
b(M, z) = 1 + \frac{1}{\delta_{c,0}} 
          \left[ \frac{\delta_c^2(z)}{\sigma^2(M)} - 1 \right] .
\label{eqn:biasmw}
\end{equation}
The effective bias factor for the host halo population is then the 
integral of the bias factor for all masses greater than $\mmin$ and 
weighted by the halo number density and lifetime: 
\begin{equation}
\beff(\mmin,z) = \left(\int_{\mmin}^{\infty} dM\;
   \frac{b(M, z) n(M, z)}{t_H(M, z)}\right)
    \left(\int_{\mmin}^{\infty} dM\;\frac{n(M, z)}{t_H(M, z)}\right)^{-1}.
\label{eqn:beff}
\end{equation}

The clustering amplitude of QSOs can be parametrized as the radius of 
a top-hat sphere in which the rms fluctuations of QSO number counts 
$\sigma_Q$ is unity:
\begin{equation}
\sigma_Q(r_1, z) = \beff(\mmin, z) \, \sigma(r_1) \, D(z) = 1, 
\end{equation}
where $\sigma(r_1)$ is the rms linear mass fluctuation in spheres of radius 
$r_1$ at $z =0$  and $D(z)$ is the linear growth factor. 
The radius $r_1$ is similar to the correlation length $r_0$, at which the 
correlation function $\xi(r)$ is unity, but can be determined more 
robustly because it is an integrated quantity and does not require 
fitting $\xi(r)$. 

  \begin{figure}
    \centering
%    \vspace{1cm}
    \includegraphics[height=110mm,width=110mm]{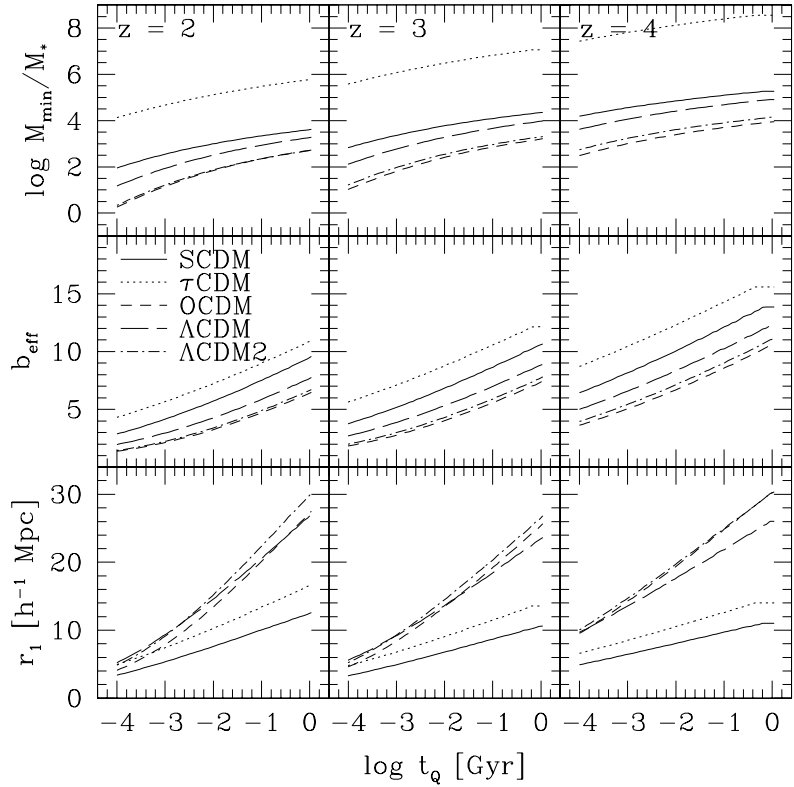}
    \caption{Minimum halo mass, effective bias, and clustering length 
as a function of $\tq$ for a range of CDM models at $z = 2, 3,$ and $4$ 
(from Martini \& Weinberg 2001; their Fig.~7). }
    \label{fig:r1a}
  \end{figure}

For some specified cosmology and comoving QSO space density, $r_1$ 
depends directly on $\tq$. Figure~\ref{fig:r1a} shows the minimum 
mass, effective bias factor, and $r_1$ as a function of $\tq$ for 
a range of cosmologies at $z = 2, 3,$ and $4$. The bottom panels, 
which show the relationship between $r_1$ and $\tq$, demonstrate that 
the clustering length increases as a function of QSO lifetime and that 
the clustering length can be used to obtain a good estimate of the 
lifetime if the cosmological parameters are relatively well known. 
The difference in the relationship between $r_1$ and $\tq$ as a function 
of redshift mostly reflects the evolution in the QSO space density (taken 
from Warren, Hewett, \& Osmer 1994). QSOs are rarer at $z = 4$ than at $z = 2$ 
or $3$ and therefore the observed space density can be matched with a 
larger $\mmin$ at the same $\tq$. At all of these redshifts 
the most massive halos correspond to the masses of individual galaxy 
halos. At lower redshifts, the most massive halos could contain 
multiple galaxies and the assumption of one QSO per halo may break down. 
The relation between clustering and lifetime is much shallower for the 
$\Omega_M = 1$ models because of their smaller mass fluctuations. For
these models the values of $\mmin$ also lie out on the steep, high-mass 
tail of the halo mass function, where a smaller change in $\mmin$ 
is required to compensate for the same change in $\tq$. 

The relationship between lifetime and clustering is sensitive to 
several model details, which can be used to either determine the 
precision of the estimated lifetime or observationally test and 
further refine the relation. One uncertainty lies in the definition 
of the halo lifetime, which was taken to be the median time before the 
halo was incorporated into a new halo of twice the original mass. If 
the definition were increased to a factor of five increase in mass 
(a strong upper limit), then the QSO lifetimes corresponding to the same 
observed $r_1$ would increase by a factor of $2 - 4$. Another 
potential complication is that the correlation between halo mass and 
QSO luminosity is not perfect. Scatter in this relation will lead to 
the inclusion of a larger number of less massive halos in an 
absolute-magnitude limited sample of QSOs. Because this scatter 
always serves to increase the number of lower-mass halos, it decreases 
the clustering signal from the ideal case of a strict relationship 
and effectively makes the clustering predictions in Figure~\ref{fig:r1a}
into lower limits for $\tq$ for a measured $r_1$. The presence of 
scatter in this relation could be determined and measured observationally. 
In the absence of scatter, there should be a direct relationship between 
the clustering length and QSO luminosity at fixed redshift. The presence 
of scatter will flatten the predicted relation between clustering and 
luminosity and could be used to estimate and correct for the amount 
of scatter in the determination of $\tq$. 

  \begin{figure}
    \centering
%    \vspace{1cm}
    \includegraphics[height=90mm,width=90mm]{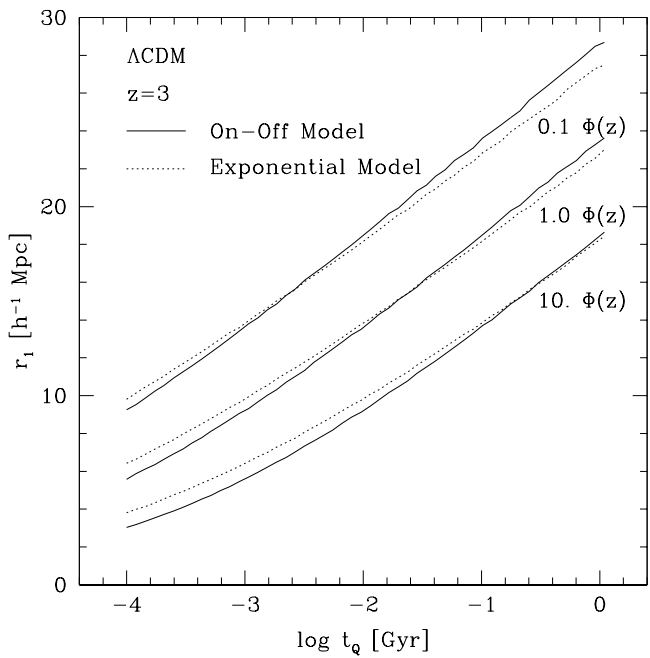}
    \caption{Clustering length vs. $\tq$ for the $\Lambda$CDM model for 
two luminosity evolution models and three values of the space density 
(from Martini \& Weinberg 2001; their Fig.~8). }
    \label{fig:r1b}
  \end{figure}

This model also assumes that the luminosity evolution of QSOs is a 
step function, that is they are either on at some constant luminosity,  
or off. If QSOs are instead triggered by a mechanism that drives a 
great deal of fuel toward the center, such as a galaxy merger, this fuel 
supply will probably gradually diminishes over time and cause the fading of 
the QSO. An alternative parameterization of QSO luminosity evolution, 
employed by Haehnelt \etal\ (1998) is an exponential decay: 
$L(t) \propto {\rm exp} (-t/\tq)$. 
Martini \& Weinberg (2001) found that if this luminosity evolution is 
used instead of the ``on-off'' model, and $\tq$ redefined to represent the
$e$-folding time of the QSO, there is very little 
change in the relation between $\tq$ and $r_1$ (see Fig.~\ref{fig:r1b}). 
This figure also 
demonstrates the scalable nature of the model to QSO surveys with 
different absolute magnitude limits. A decrease in the absolute magnitude 
limit is equivalent to increasing the space density of QSOs and therefore 
the number of lower-mass halos. To first approximation, increasing the 
space density of the sample by a factor of 10 corresponds to a factor of 
10 longer lifetime at fixed $r_1$. These results scale in a similar manner
if QSOs are beamed. If QSOs are only visible through some beaming angle 
$f_B$, then the space density needs to be corrected by a factor $f_B^{-1}$ 
before the QSO lifetime can be determined from the clustering. 

The 2dF and SDSS surveys will provide the best measurements of QSO clustering
to date. Preliminary results from these surveys (Croom \etal\ 2001; 
Fan 2003) indicate that $r_0 \approx 6 \hmpc$, or 
$r_1 \approx 9 \hmpc$ for $\xi(r) = (r/r_0)^{-1.8}$. This corresponds to 
a lifetime of $\tq \approx 10^6$ years. A much larger correlation length 
$r_0 = 17.5 \pm 7.5 \hmpc$ (Stephens \etal\ 1997) was measured in the 
Palomar Transit Grism Survey (Schneider, Schmidt, \& Gunn 1994). 
This high correlation length may be a statistical anomaly as it is only based 
on three QSO pairs, but could also be a consequence of the very high 
luminosity threshold of the survey. 

\section{Summary} \label{sec:con}

\vspace{0.1in}

 \begin{table}
  \caption{Constraints on the QSO Lifetime}
    \begin{tabular}{l|cc}
     \hline \hline
     {Method} & {Net/Episodic} & Lifetime [yr]\\
     \hline
     {\bf Current:} & & \\
     Evolution 			& Net 		& $<10^9$ \\
     Local Black Holes 		& Net 		& $10^7 - 10^8$ \\
     Merger Scenario 		& Net 		& few $\times 10^7$ \\
     Mergers \& Starbursts 	& Episodic 	& few $\times 10^8$ \\
     Proximity Effect 		& Episodic 	& $>10^4$ \\
     Radio Jets 		& Episodic 	& few $\times 10^8$ \\	
	& & \\
     {\bf Future:} & & \\
     QSOs at $z>6$  		& Episodic 	& few $\times 10^7$? \\	
     Transverse Proximity Effect& Episodic 	& $10^7$ ?? \\
     Clustering 		& Net 		& $10^6 - 10^7$ ?? \\
     \hline \hline
    \end{tabular}
  \label{sample-table}
 \end{table}

Current estimates of the QSO lifetime lie in the range $\tq = 10^6 - 10^8$ 
years, still uncertain by several orders of magnitude.
Estimates of the net lifetime are primarily demographic and rely 
on assumptions about the radiative efficiency of accretion $\epsilon$ and 
the luminosity of QSOs relative to the Eddington rate $\lle$. The new models 
developed in the last few years (Kauffmann \& Haehnelt 2000; Yu \& 
Tremaine 2002) have started to relax the common assumptions about 
the values of $\epsilon$ and $\lle$, but a complex relationship between 
$\epsilon$, $\lle$, and $M_{\bullet}$ has by no means been ruled out. 

While the physics of the accretion process and the potential for 
significant obscured black hole growth may appear daunting obstacles, 
there are reasons for optimism. Demographic estimates of the net lifetime 
will be substantially improved as more supermassive black hole masses 
are measured and the validity of the $M_\bullet - \sigma$ relation better 
established, 
including its application to active galaxies. Measurement of the black hole 
mass in more QSOs via reverberation mapping 
(\eg, Barth 2003), or estimates using the 
$M_\bullet - \sigma$ 
method will lead to direct estimates of $\lle$ for 
many QSOs. These steps will help to eliminate some of the main uncertainties 
in the value or range of $\lle$ and the local mass density of supermassive 
black holes. 
Finally, measurement of a significant number of redshifts for 
the sources that comprise the hard X-ray and sub-mm background will greatly 
improve the integral constraint provided by the net luminosity from accretion.
Many of these new observations could be realized in the next several years 
and reduce the need for sweeping assumptions about $\epsilon$ or 
$\lle$. 

The current best estimates or limits for the episodic lifetime are consistent 
with a wider range of $\tq$ than the net lifetime, but constraints from the 
proximity effect in the \lya\ forest do point to $\tq > 10^4$ years. 
Observations of QSOs when the opacities of neutral helium and hydrogen was 
much higher suggest much longer episodic lifetimes, 
although to date are still only based on observations of two QSOs. 
Constraints such as these on the episodic lifetime provide a valuable 
complement to the primarily demographic estimates of the net lifetime, as they 
do not depend on assumptions of accretion physics, but instead on much 
more straightforward radiation physics. 

The next few years offer the hope of tremendous progress in measurement of 
the QSO lifetime via a variety of techniques. 
The discovery of additional QSOs associated with He~II absorption systems 
(and hopefully more of these systems), as well as additional QSOs 
at high redshift $z > 6$, could provide quite strong lower limits to 
the episodic lifetime, while clustering measurements hold great promise to 
constrain the net lifetime. As the clustering method does not depend on 
assumptions about accretion physics, it will provide a valuable complement 
to the present demographic approaches. 

This potential progress suggests that $\tq$ could be determined to within a
factor of 3 in the next three years, which is sufficient precision to  
address some fundamental questions on the growth of black holes and the 
physics of the accretion process. One important advance will be to 
determine if the QSO lifetime determined via clustering is 
in agreement with that estimated via black hole demographics. As these 
models employ an independent set of assumptions, these measurements could 
provide strong evidence that $\epsilon \approx 0.1$ and $\lle \approx 1$ 
are valid choices. 

Measurement of $\tq$ will also determine if a luminous QSO phase was the 
dominant growth mechanism for present-day supermassive black holes. 
If $\tq > 4 \times 10^7$ yr, then a substantial fraction of the black hole 
mass in QSO hosts was accreted via this optically luminous mechanism. 
However, such a long lifetime would also imply that QSOs were quite rare 
phenomena
and only a small fraction of present-day supermassive black holes were 
created as QSOs. The remainder may have been identifiable as AGN, 
or were heavily obscured, but would not have been classified as optically 
luminous QSOs. 

If instead $\tq$ is shorter than the Salpeter time scale, then hosting  
a short-lived QSO phase was a relatively common occurrence. However, only 
a small fraction of the present-day black hole mass was accreted during this 
time and the remainder must have occurred via a less luminous, or less 
radiatively efficient, mode of accretion, either before or after the 
QSO epoch. This could then imply that 
advection-dominated accretion (\eg, Narayan, Mahadevan, \& Quataert 1998) or 
inflow-outflow solutions (Blandford \& Begelman 1999) played a 
significant role in black hole growth. 

\vspace{0.1in}

I would like to thank Zolt\'an Haiman and David Weinberg for helpful 
comments on this manuscript. I am also grateful to Luis Ho for 
inviting and encouraging me to prepare this review.  

%\section{References}


\begin{thereferences}{}

\vspace{0.1in}

\bibitem{}
Bajtlik, S., Duncan, R.~C., \& Ostriker, J.~P. 1988, \apj, 327, 570 

\bibitem{}
Barger, A.~J., Cowie, L.~L., Bautz, M.~W., Brandt, W.~N., Garmire, G.~P.,
Hornschemeier, A.~E., Ivison, R.~J., \& Owen, F.~N. 2001, \aj, 122, 2177

\bibitem{}
Barth, A.~J. 2003, in 
Carnegie Observatories Astrophysics Series, Vol. 1: Coevolution of Black
Holes and Galaxies, ed. L. C. Ho (Pasadena: Carnegie Observatories,
http://www.ociw.edu/ociw/symposia/series/symposium1/proceedings.html)

\bibitem{}
Begelman, M.~C. 2002, \apj, 568, L97 

\bibitem{}
Bennert, N., Falcke, H., Schulz, H., Wilson, A.~S., \& Wills, B.~J. 2002, 
\apj, 574, L105

\bibitem{}
Bernardi, M., \etal\ 2003, \aj, 125, 1817

\bibitem{}
Blandford, R.~D., \& Begelman, M.~C. 1999, \mnras, 303, L1

\bibitem{}
Blundell, K.~M., Rawlings, S., \& Willott, C.~J. 1999, \aj, 117, 677

\bibitem{}
Boyle, B.~J, Shanks, T., Croom, S.~M., Smith, R.~J., Miller, L., Loaring,
N., \& Heymans, C. 2000, \mnras, 317, 1014

\bibitem{}
Brotherton, M.~S., et al.  1999, \apj, 520, L87

\bibitem{}
Canalizo, G., \& Stockton, A. 2001, \apj, 555, 719 

\bibitem{}
Cavaliere, A., \& Padovani, P. 1988, \apj, 333, L33

\bibitem{}
------. 1989, \apj, 340, L5

\bibitem{}
Cen, R., \& McDonald, P. 2002, \apj, 570, 457

\bibitem{}
Croom, S.~M., Boyle, B.~J., Loaring, N.~S., Miller, L., Outram, P.~J.,
Shanks, T., \& Smith, R.~J. 2002, \mnras, 335, 459

\bibitem{}
Elvis, M., \etal\ 1994, \apjs, 95, 1

\bibitem{}
Fan, X. 2003, in Carnegie Observatories Astrophysics Series, Vol. 1: 
Coevolution of Black Holes and Galaxies, ed. L. C. Ho (Pasadena: Carnegie 
Observatories, 
http://www.ociw.edu/ociw/symposia/series/symposium1/proceedings.html)

\bibitem{}
Ferrarese, L., \& Merritt, D. 2000, \apj, 539, L9

\bibitem{}
Gebhardt, K., \etal\ 2000, \apj, 539, L13 

\bibitem{}
Haehnelt, M.. \& Natarajan, P., \& Rees, M.~J. 1998, \mnras, 300, 817

\bibitem{}
Haehnelt, M., \& Rees, M.~J. 1993, \mnras, 263, 168 

\bibitem{}
Haiman, Z., \& Cen, R. 2002, \apj, 578, 702 

\bibitem{}
Haiman, Z., \& Hui, L. 2001, \apj, 547, 27 

\bibitem{}
Haiman, Z., \& Loeb, A. 1998, \apj, 503, 505

\bibitem{}
------. 2001, \apj, 552, 459

\bibitem{}
Heap, S.~R., Williger, G.~M., Smette, A., Hubeny, I., Sahu, M., Jenkins,
E.~B., Tripp, T.~M., \& Winkler, J.~N. 2000, \apj, 534, 69

\bibitem{}
Hewett, P.~C., Foltz, C.~B.,, \& Chaffee, F.~H. 1995, \aj, 109, 1498 

\bibitem{}
Jakobsen, P., Jansen, R.~A., Wagner, S., \& Reimers, D. 2003, \aa, 397, 891

\bibitem{}
Kauffmann, G., \& Haehnelt, M. 2000, \mnras, 311, 576 

\bibitem{}
Kawakatu, N., Umemura, M., \& Mori, M. 2003, in Carnegie Observatories 
Astrophysics Series, Vol. 1: Coevolution of Black Holes and Galaxies, ed. 
L. C. Ho (Pasadena: Carnegie Observatories, 
http://www.ociw.edu/ociw/symposia/series/symposium1/proceedings.html)

\bibitem{}
Longair, M.~S., \& Riley, J.~M. 1979, \mnras, 188, 625

\bibitem{}
Lynden-Bell, D. 1969, Nature, 223, 690

\bibitem{}
Martini, P., \& Weinberg, D.~H. 2001, \apj, 547, 12

\bibitem{}
Mo, H.~J., \& White, S.~D.~M. 1996, \mnras, 282, 347 

\bibitem{}
Narayan, R., Mahadevan, R., \& Quataert, E. 1998, in The Theory of Black Hole
Accretion Discs, ed.  M. A. Abramowicz, G. Bj\"{o}rnsson, \& J. E. Pringle
(Cambridge: Cambridge Univ. Press), 148

\bibitem{}
Norman, C.~A., \& Scoville, N.~Z. 1988, \apj, 332, 124

\bibitem{}
Osmer, P.~S. 2003, in Carnegie Observatories Astrophysics Series, Vol. 1: 
Coevolution of Black Holes and Galaxies, ed. L. C. Ho (Cambridge: 
Cambridge Univ. Press) 

\bibitem{}
Press, W.~H., \& Schechter, P. 1974, \apj, 187, 425 

\bibitem{}
Rees, M.~J. 1984, \araa, 22, 471 

\bibitem{}
Richstone, D.O., \etal\ 1998, \nature, 395, A14

\bibitem{}
Salpeter, E.~E. 1964, \apj, 140, 796 

\bibitem{}
Salucci, P., Szuszkiewicz, E., Monaco, P., \& Danese, L. 1999, \mnras, 307, 637

\bibitem{}
Scheuer, P.~A.~G. 1995, \mnras, 277, 331 

\bibitem{}
Schneider, D.~P., Schmidt, M., \& Gunn, J.~E. 1994, \aj, 107, 1245

\bibitem{}
Scott, J., Bechtold, J., Dobrzycki, A., \& Kulkarni, V.~P. 2000, \apjs, 130, 67

\bibitem{}
Small, T.~A., \& Blandford, R.~D. 1992, \mnras, 259, 725 

\bibitem{}
So\l tan, A. 1982, \mnras, 200, 115

\bibitem{}
Stephens, A.~W., Schneider, D.~P., Schmidt, M., Gunn, J.~E., \& Weinberg, D.~H.
1997, \aj, 114, 41

\bibitem{}
Thorne, K.~S. 1974, \apj, 191, 507 

\bibitem{}
Warren, S.~J., Hewett, P.~C., \& Osmer, P.~S. 1994, \apj, 421, 412

\bibitem{}
Yu, Q. 2003, in 
Carnegie Observatories Astrophysics Series, Vol. 1: Coevolution of Black 
Holes and Galaxies, ed. L. C. Ho (Pasadena: Carnegie Observatories, 
http://www.ociw.edu/ociw/symposia/series/symposium1/proceedings.html)

\bibitem{}
Yu, Q., \& Tremaine, S. 2002, \mnras, 335, 965
 
\bibitem{}
Zel'dovich, Ya.~B., \& Novikov, I.~D. 1964, Sov. Phys. Dokl., 158, 811

\end{thereferences}
\end{document}